\documentclass[preprint,showpacs,superscriptaddress]{revtex4-1}

\usepackage{graphicx}
\usepackage{epstopdf}

\begin{document}


\title{Preventing rapid energy loss from electron-hole pairs to phonons in graphene quantum dots}


\author {J.P. Trinastic}
 \affiliation {Department of Physics and Quantum Theory Project, University of Florida, Gainesville, Florida, 32611, USA}
\author {Iek-Heng Chu}
\author {Hai-Ping Cheng*}
 \affiliation {Department of Physics and Quantum Theory Project, University of Florida, Gainesville, Florida, 32611, USA}


\date{\today}




\maketitle



\textbf{In semiconductors, photoexcited electrons and holes (carriers) initially occupy high-energy states, but quickly lose energy to phonons and relax to the band edge within a picosecond \cite{PhysRevLett.72.1364}.  Increasing the lifetime of carriers in light-absorbing materials is necessary to improve open-circuit voltage in photovoltaics \cite{conibeer2008slowing}, charge separation in organic solar cells \cite{jailaubekov2013hot}, and charge transfer in photodetection devices \cite{doi:10.1021/jz900022z}.  Here we demonstrate long lifetimes over one hundred picoseconds for electron-hole pairs in graphene quantum dots (GQDs) due to large transition energies and weak coupling to excitonic states below the fundamental band gap.  This possibility for a large transition energy to bound excitons is due to graphene's poor screening, illustrating a unique mechanism in this QD to occupy higher-energy states for long timescales.  GQD edges can be terminated with either armchair or zigzag carbon patterns, and this edge structure changes excited state  lifetimes by orders of magnitude.  These results indicate nanoscale control of carrier lifetimes in optoelectronics.}

Semiconductor quantum dots (QDs) exhibit discrete transition energies between excited states due to electronic confinement at the dot boundary.  In optoelectronic applications, a photoexcited carrier will transition to lower-energy excited states due to energy loss to lattice phonons, known as phonon-induced relaxation (slow relaxation: long state lifetime; fast relaxation: short lifetime).  Energy must be conserved during this process such that the total energy of coupled phonons matches the transition energies around 0.1 eV \cite{Pandey07112008}.  This transition energy often mismatches available phonon energies, typically tens of meV \cite{peterson2014role}, so that multiple phonons must facilitate the transition to lower energies, leading to inefficient carrier relaxation that should create long excited state carrier lifetimes in QDs ("phonon bottleneck").  Electronic coupling between involved states must also be weak, which depends on the wave functions and the momentum of phonon modes inducing the transition \cite{prezhdo1997evaluation}.

Understanding how to engineer phonon bottlenecks can improve optoelectronic devices \cite{conibeer2008slowing,jailaubekov2013hot}, however the promise of long excited carrier lifetimes beyond tens of picoseconds (ps) has not been realized experimentally in QDs due to defects and surface ligands \cite{guyot2005intraband,PhysRevLett.80.4028,PhysRevLett.98.177403}.  These structural factors introduce additional states that reduce transition energies and increase electron-phonon coupling, leading to shortened carrier lifetimes in the picosecond range \cite{guyot2005intraband}.  Thus, finding QDs with a phonon bottleneck and determining how structural changes control its timescale are priorities for improving optoelectronic performance.

In this Letter, we demonstrate a unique method to create a phonon bottleneck in GQDs by taking advantage of graphene's low dielectric constant \cite{hwang2012fermi}.  The Coulomb interaction is not effectively screened in graphene and a strong attraction between excited electrons and holes creates bound states, known as excitons, with energies significantly below the fundamental band gap.  Originally described in 1931 by Frenkel, the exciton is a fundamental many-body excitation with neutral charge, consisting of an electron-hole pair typically localized within one molecule or unit cell.  Manipulating excitons has provided insights into charge transfer states at organic interfaces \cite{deibel2010role}, optical gain and laser emission \cite{PhysRevLett.69.1707}, and the excited state properties of low-dimensional materials \cite{andreani1995optical}.  The low screening in GQDs should create a large transition energy between higher-energy states and excitons, requiring multiple phonons to induce a transition.  This could lead to long excited state carrier lifetimes if other excited states have weak electronic coupling to excitons.  GQDs are synthesized with chemical methods that minimize defects and allow for nanoscale structural control \cite{yan2010large} that we show can tune this electronic coupling.

\begin{figure}[h]
    \includegraphics[scale=1]{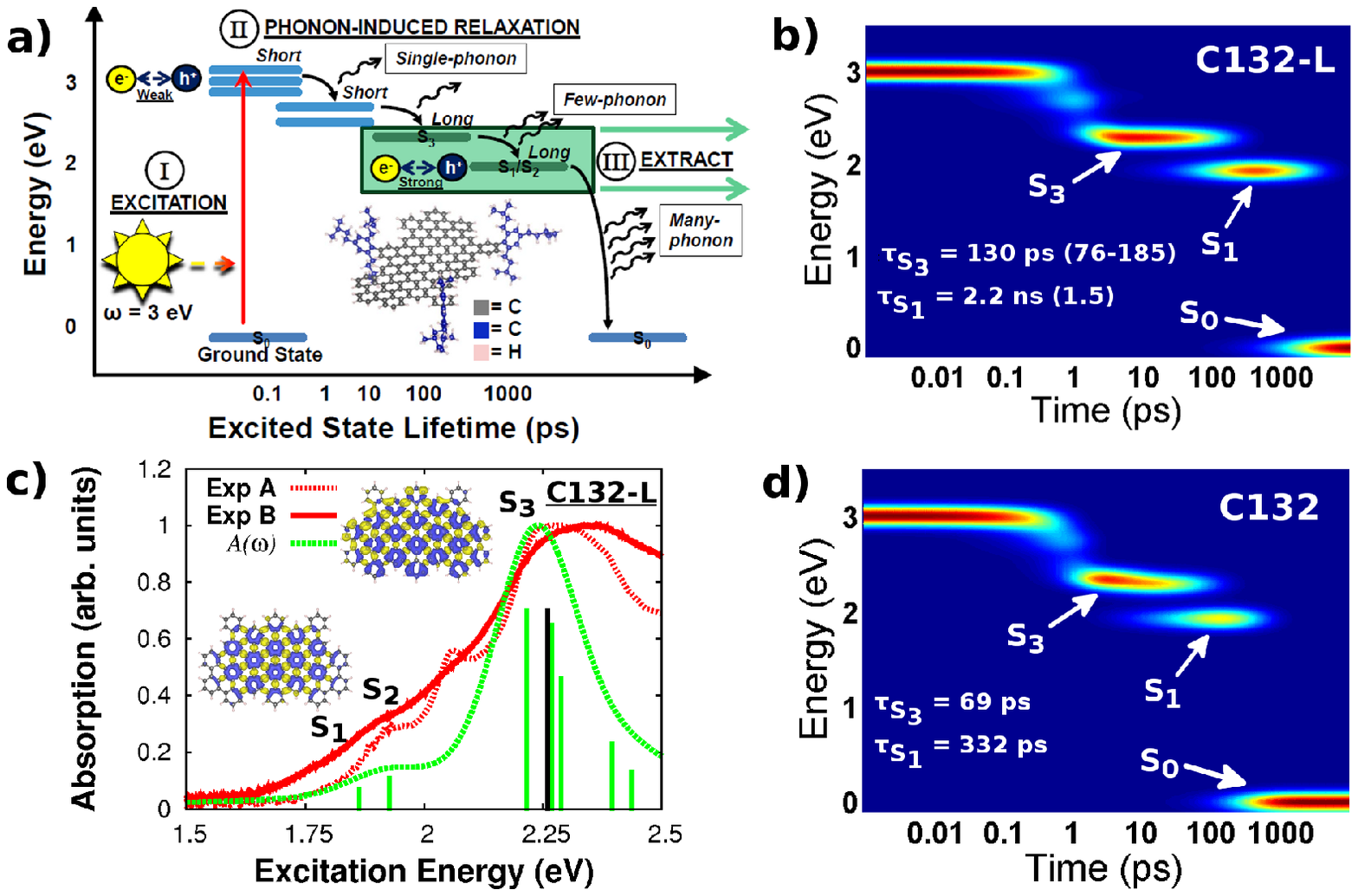}
    \caption{Summary of phonon-induced relaxation process and optical absorption for 132-carbon-atom graphene quantum dot (GQD). a) Diagram illustrating physical processes relevant to our model (see text for details). b) Model predictions of the energy as a function of time and state lifetimes ($\tau$) of an excited carrier in C132-L after 3 eV photoexcitation.  Colors from blue to red indicate population changes from 0 to 1, where 1 represents the entire population in one state. Experimental lifetimes taken from transient absorption measurements \cite{mueller_hot_carrier} are included in parentheses.  c) Normalized absorption spectra of C132-L GQD comparing experimental (Exp A \cite{mueller_hot_carrier} and B \cite{williams_gqd_tio2}) to calculated results ($A(\omega$)).  Vertical lines indicate the calculated excitation energies of the system (heights do not indicate a physical quantity).  The black line indicates the HOMO-LUMO electronic band gap.  Yellow and blue isosurfaces on the GQD structures indicate electron and hole charge densities of each excitation, emphasizing the localized nature of S$_1$ and S$_2$ (near-indentical charge densities) near the center of the GQD compared to higher excited states.  d) Population relaxation dynamics for GQD without ligands (C132) similar to (b).}
  \label{comp_sc_uc}
\end{figure}

We model phonon-induced relaxation of electron-hole pairs in GQDs using a state-of-the art reduced density matrix method that rigorously accounts for excitons (see Methods).  We first consider a 132-carbon-atom GQD with carbon ligands (C132-L) studied experimentally \cite{mueller_hot_carrier}.  Figure \ref{comp_sc_uc}(a) shows the general excitation and phonon-induced relaxation process predicted by our model results in Figure \ref{comp_sc_uc}(b).  To match transient absorption (TA) measurements \cite{mueller_hot_carrier}, we initially excite an electron-hole pair with a 3 eV photoexcitation (I), creating a "hot" electron-hole pair weakly bound by their Coulomb attraction and delocalized across the edges of the GQD.  Over time, the pair loses energy to phonons (II), rapidly transitioning downward through single-phonon processes over the first 0.1-1 ps.  Within 10 ps, the electron-hole pair transitions to a state near 2.25 eV (S$_3$ in Figure \ref{comp_sc_uc}(a)-(c)) that corresponds to the absorption peak of the molecule near the fundamental band gap (Figure \ref{comp_sc_uc}(c)).

The S$_3$ state, however, does not correspond to the lowest excited state in GQDs, since poor screening leads to two excitonic states (S$_1$ and S$_2$ in Figure \ref{comp_sc_uc}(c)) 0.30 eV below the absorption peak that match the experimental rise in absorption near 1.85 eV (red curves in Figure \ref{comp_sc_uc}(c); Exp A has a small peak at 2.05 eV to be explored in future work).  The large transition energy to the excitons necessitates a few-phonon process that drastically slows carrier relaxation, giving S$_3$ a long 130 ps lifetime before transitioning to S$_1$/S$_2$, localized near the center of the dot.  This 100 ps timescale is 1-2 orders of magnitude longer than relaxation in bulk graphene \cite{dawlaty_graph_rel} and an order of magnitude longer than lifetimes in other QDs \cite{kilina2008breaking,guyot2005intraband}.  The bound electron-hole pair stays in the excitonic states for two nanoseconds before nonradiatively relaxing to the ground state through a many-phonon process.  Both the S$_3$ (130 ps) and S$_1$ (2.2 ns) lifetimes match extremely well with 76-185 ps and 1.5 ns \cite{mueller_hot_carrier} lifetimes observed at similar energies in TA measurements.  The 100 ps lifetime provides the potential to extract excited carriers from both the higher-energy excited states and excitonic states (Figure \ref{comp_sc_uc}(a), green region) in photovoltaic or other applications, as extraction times under 100 ps are possible to acceptor materials such as titanium oxide \cite{Tisdale18062010,jacs_cdse_tio2_inject}.  To our knowledge, these are the first $ab$ $initio$ results demonstrating slow carrier relaxation to excitons in a poor-screening material like graphene.

\begin{figure}[h]
    \includegraphics[scale=1]{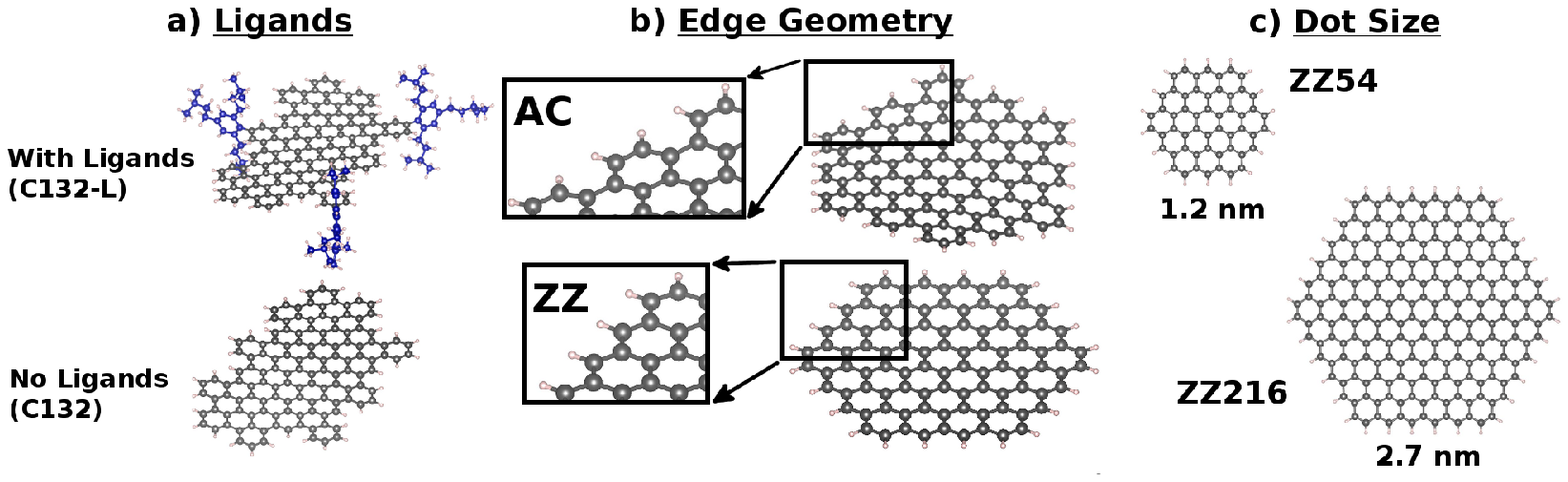}
    \caption{Structural modifications to GQDs investigated to determine their effect on excited carrier lifetimes. a) 132-carbon-atom GQD with (C132-L) and without (C132) carbon-chain ligands used in experiment to prevent dot aggregation. b) GQDs with carbon atom edge termination following an armchair (AC) or zigzag (ZZ) pattern.  c) GQDs of increasing size with diameters ranging from 1 to 3 nanometers.  GQDs are labeled by their edge type (AC or ZZ) and number of carbon atoms.  For example, ZZ216 refers to a GQD with 216 carbon atoms and zigzag termination edges.}
  \label{struc_mod}
\end{figure}

Finding ways to modify state lifetimes would provide a path to tailor relaxation dynamics for various optoelectronic devices.  Therefore, we next show how nanoscale structural changes can tune excited carrier lifetimes.  As shown in Figure \ref{struc_mod}, we explore three structural modifications: a) removing the carbon-chain ligands on the experimental C132-L GQD (C132), b) changing the carbon edge geometry terminating the GQD between armchair (AC) and zigzag (ZZ) patterns, and c) increasing the GQD size from 1-3 nanometers.  For example, ZZ216 refers to a GQD with 216 carbon atoms and zigzag termination edges.

Looking at the effect of ligands, Figures \ref{comp_sc_uc}(b) and \ref{comp_sc_uc}(d) plot relaxation dynamics for C132-L and C132, respectively.  Lifetimes at high-energy states are similar, however the S$_3$ lifetime ($\tau_{S_3}$) is shortened by half in C132 (69 ps) compared to C132-L (130 ps).  More dramatically, the S$_1$ lifetime ($\tau_{S_1}$) decreases by an order of magnitude from 2.2 ns to 332 ps in C132, indicating that ligands delay nonradiative recombination to the ground state and could increase fluorescence yields.  Previous work has indicated ligands expedite relaxation by introducing defects or hybridized states \cite{kilina2012surface}, however we find that ligands increase lifetimes through physical mechanisms we explore below.

\begin{figure}[h]
    \includegraphics[scale=1]{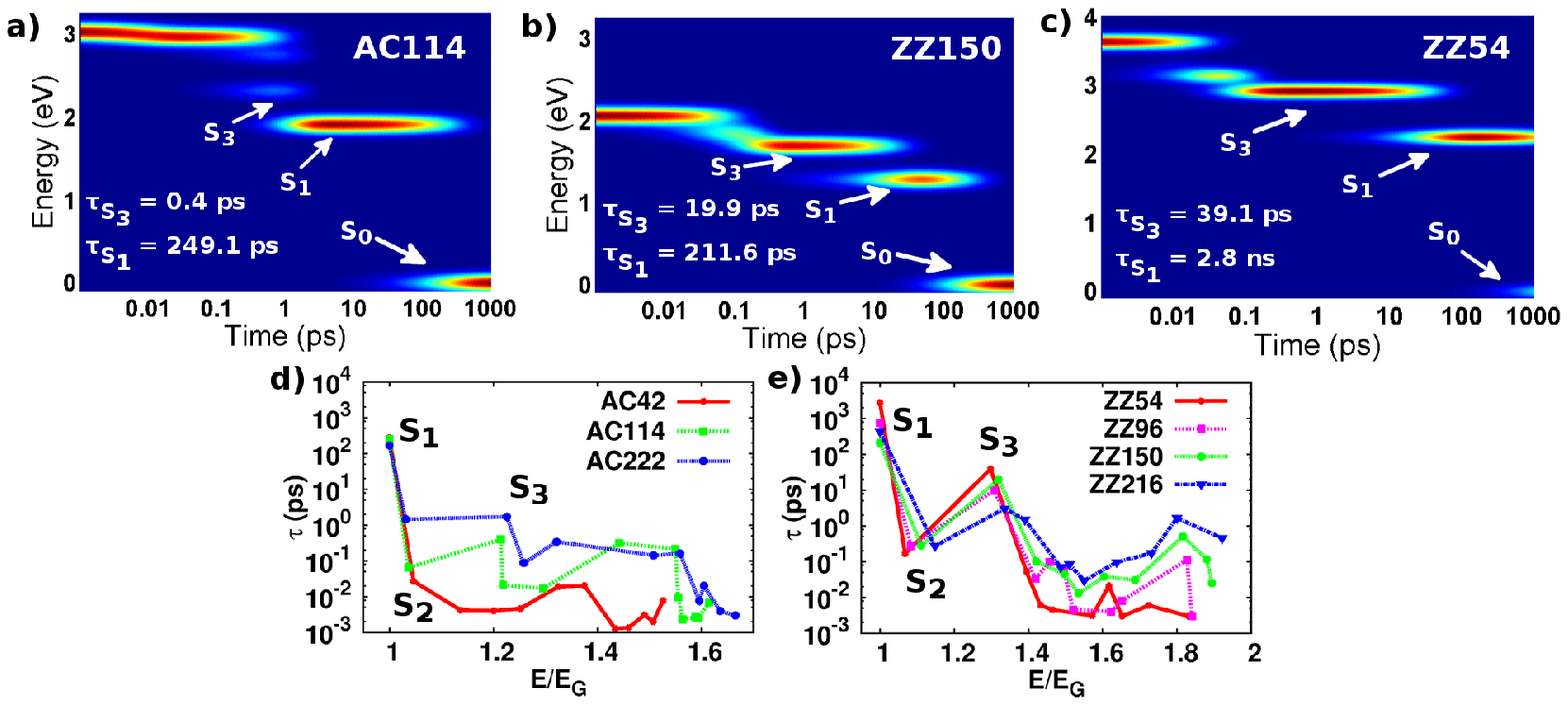}
    \caption{Summary of relaxation dynamics for GQDs with armchair (AC) and zigzag (ZZ) termination edges.  All phonon-induced relaxation occurs after a 1.6 $E/E_G$ photoexcitation, where $E_G$ is the energy of the lowest excitation.  Colors from blue to red indicate population changes from 0 to 1, where 1 represents the entire population in one state at a given energy and time. Figures a) through c) plot the energy of the excited carrier as a function of time for AC114, ZZ150, and ZZ54, respectively, and list lifetimes ($\tau$) of important states.  In all cases, S$_3$ corresponds to the absorption peak close to the electronic band gap and S$_1$ is an excitonic state.  d) and e) plot carrier lifetimes ($\tau$) as a function of $E/E_G$ and size for AC and ZZ GQDs, respectively.}
  \label{ac_zz_rel}
\end{figure}

Modifying carbon edge type (Figure \ref{struc_mod}(b)) and size (Figure \ref{struc_mod}(c)) reveals a dramatic ability to tune the phonon bottleneck to excitonic states.  All GQDs demonstrate an absorption peak at S$_3$ and two excitonic states (S$_1$ and S$_2$) 0.23-0.53 eV below the fundamental band gap, with smaller GQDs exhibiting larger transition energies (see Tables SI-SII).  Comparing GQDs similar in size, we plot excited carrier energy as a function of time for AC114 and ZZ150  in Figures \ref{ac_zz_rel}(a)-(b).  In both cases, the excited carrier relaxes to the absorption peak (S$_3$) under a picosecond.  However, the S$_3$ lifetime for AC114 ($\tau^{AC114}_{S_3}$) dramatically reduces to 0.4 ps, breaking the phonon bottleneck and providing a fast relaxation channel to excitons.  Across all AC sizes (Figure \ref{ac_zz_rel}(d)), S$_3$ lifetimes are never more than 2 ps and higher-energy state lifetimes are within 10-50 fs ($E/E_G >$ 1.5, where $E_G$ is the first excitation energy).  In contrast, the S$_3$ lifetime is two orders of magnitude larger for ZZ150 ($\tau^{ZZ150}_{S_3} =$ 19.9 ps) and further lengthened by decreasing ZZ GQD size ($\tau^{ZZ54}_{S_3} =$ 39.1 ps, Figure \ref{ac_zz_rel}(c)).  Increasing ZZ GQD size leads to a sharp decrease in S$_3$ lifetime to 3.1 ps for ZZ216 (Figure \ref{ac_zz_rel}(e)).

An intriguing effect of size on lifetimes occurs for high-energy states in ZZ GQDs (Figure \ref{ac_zz_rel}(e)).  Wider energy spacing at $E/E_G >$ 1.7 (Figure S1) leads to increased lifetimes as a function of size, reaching picosecond lifetimes in ZZ216 (Figure \ref{ac_zz_rel}(e)).  Multiexciton generation (MEG) can occur in states with $E/E_G >$ 2 if phonon-induced relaxation is slow enough to allow high-energy carriers to Coulombically interact with another electron-hole pair and decay into multiple low-energy carriers.  This process leads to multiple carriers per absorbed photon and increased photovoltaic efficiency.  MEG occurs on timescales of 100 fs to 100 ps in CdSe \cite{rabani2008distribution} and 250 fs in PbSe or PbS \cite{ellingson2005highly} QDs, competitive with the high-energy lifetimes seen in larger ZZ GQDs, encouraging future study of MEG in these systems.

The above results suggest that the phonon bottleneck to excitons has a strong sensitivity to nanoscale changes in structure.  To explore the physical mechanisms behind these differences, we consider the factors influencing transition rates in our model (see Methods and Supplementary Materials).  Phonon-induced transition rates are expressed as \cite{tokmakoff_redfield,may2008charge}: 
\begin{equation}
\label{kab}
k_{\alpha\beta}=\frac{2\pi}{\hbar}|V_{\alpha\beta}|^2 F(\omega_{\alpha\beta}),
\end{equation}

\noindent where $\omega_{\alpha\beta}$ is the transition energy between states $\alpha$ and $\beta$, $V_{\alpha\beta}$ is the electronic coupling \cite{prezhdo1997evaluation}, which measures wave function coupling via phonon momenta calculated using quantum molecular dynamics (QMD), and $F(\omega_{\alpha\beta})$ is the Franck-Condon-weighted density of states \cite{may2008charge}.  $F(\omega_{\alpha\beta})$ depends on the Huang-Rhys factor ($S_{\alpha\beta}$), which measures the coupling strength of a given mode to a transition.  The function $F(\omega)$ peaks at energies corresponding to modes with nonzero $S_{\alpha\beta}$.  Examining $V_{\alpha\beta}$ and $S_{\alpha\beta}$ as a function of phonon frequency will provide physical insight into the carrier lifetimes discussed above.

\begin{figure}[h]
    \includegraphics[scale=1]{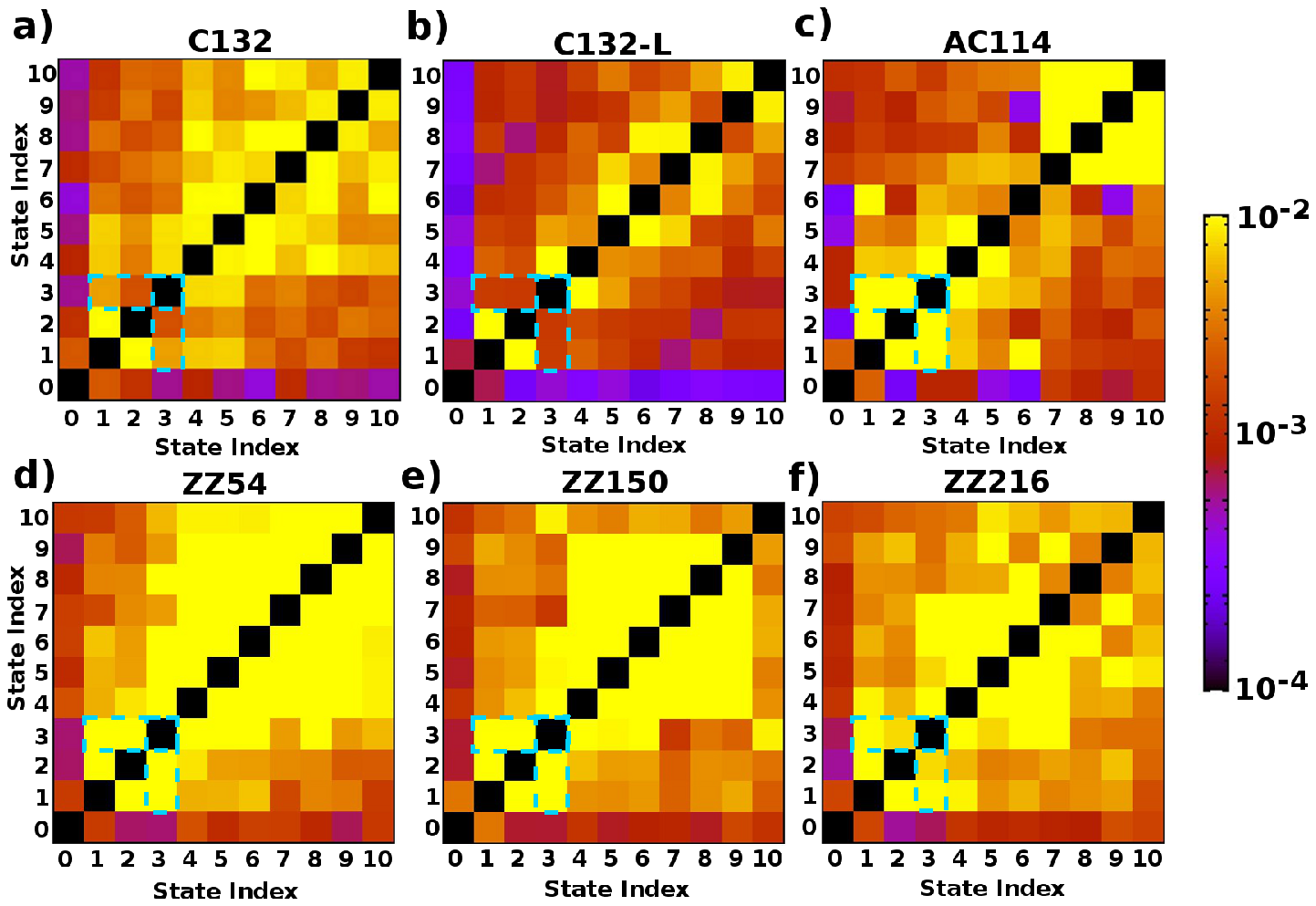}
    \caption{Electronic nonadiabatic coupling ($V_{\alpha\beta}$) for GQDs with different structural modifications. (a)-(f) plot $V_{\alpha\beta}$ matrix elements for the ground state (state index 0) and first ten excited states (state indices 1-10) for the C132, C132-L, AC114, ZZ54, ZZ150, and ZZ216 GQDs, respectively.  The matrix elements corresponding to coupling between states at the absorption peak, S$_3$, and excitonic states, S$_2$ ($V_{32}$) and S$_1$ ($V_{31}$), are highlighted by the dashed blue line.}
  \label{rates}
\end{figure}

We first consider the effect of ligands on the above factors.  Figures \ref{rates}(a)-(b) plot $V_{\alpha\beta}$ for the ground state and lowest 10 excited states (indexed 0-10) in C132 and C132-L, respectively.  $V_{\alpha\beta}$ is smaller for C132-L across most matrix elements, indicating that the ligands suppress electronic coupling.  In particular, ligands weaken coupling to excitons ($V^{C132}_{32}$ = 2.3 meV, $V^{C132-L}_{32}$ = 1.7 meV) and to the ground state ($V^{C132}_{10}$ = 2.5 meV, $V^{C132-L}_{10}$ = 0.8 meV), thus reducing $k_{\alpha\beta}$ and rationalizing the longer S$_3$ and S$_1$ lifetimes in C132-L compared to C132 (Figures \ref{comp_sc_uc}(b) and \ref{comp_sc_uc}(d)).  The reduced $V_{\alpha\beta}$  likely occurs because ligands stabilize the dot from motion perpendicular to the graphene plane (Figure S2).  During QMD, edge C atoms in C132 oscillate by several Angstroms whereas ligands in C132-L limit this motion.  Such substantial nuclear motion in C132 increases electronic coupling between states.  

\begin{figure}[h]
    \includegraphics[scale=2]{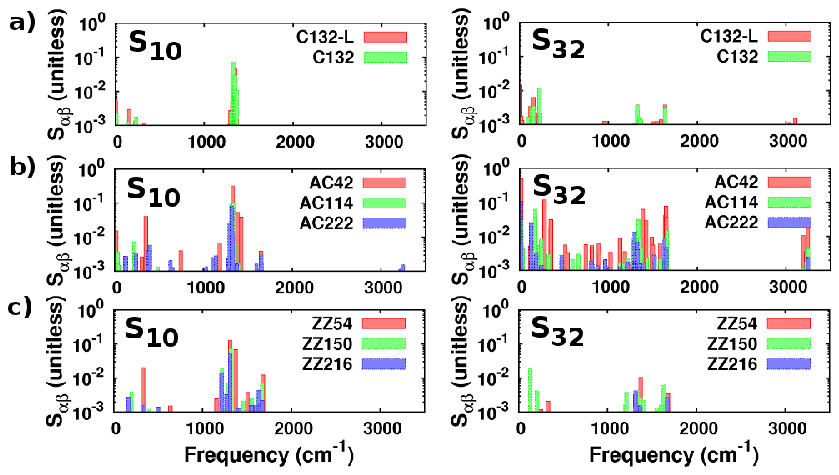}
    \caption{Huang-Rhys factors ($S_{\alpha\beta}$) as a function of phonon frequency.  a) C132-L and C132, b) AC, and c) ZZ GQDs of varying sizes.  The left and right columns refer to Huang-Rhys factors for modes coupling to the S$_1$-S$_0$ and S$_3$-S$_2$ transitions, respectively.}
  \label{hrf}
\end{figure}

Both C132 and C132-L couple to low-frequency breathing modes as well as 1300 cm$^{-1}$ (0.16 eV) and 1600 cm$^{-1}$ (0.20 eV) defect and optical G modes typical in graphene (Figure \ref{hrf}(a)) \cite{casiraghi2009raman} .  The weak coupling of the S$_3$-S$_2$ transition to these modes ($S_{32} <$ 0.01), combined with the 0.30 eV transition energy to excitons, leads to a small $F(\omega_{\alpha\beta})$ and inefficient few-phonon relaxation process that results in the 100 ps carrier S$_3$ lifetimes and phonon bottleneck.  The ligands in C132-L introduce additional phonon modes at 1000 cm$^{-1}$ and 3000 cm$^{-1}$ , however they only weakly couple to low-energy transitions and therefore do not significantly impact lifetimes (Figure \ref{hrf}(a)).  These results, combined with the reduced electronic coupling, provide a rationale for the 100 ps lifetimes seen experimentally \cite{mueller_hot_carrier}.

Examining both $V_{\alpha\beta}$ and $S_{\alpha\beta}$ also explains different relaxation patterns in AC and ZZ GQDs.  We focus on $V_{32}$ and $V_{31}$, highlighted by the dashed blue line in Figures \ref{rates}(a)-(f), as these couplings determine relaxation rates from the absorption peak (S$_3$) to excitons (S$_1$ and S$_2$).  Both AC and ZZ GQDs exhibit strong electronic coupling to excitons ($V_{\alpha\beta} >$ 10 meV).  Whereas $V_{\alpha\beta}$ in AC GQDs does not significantly change with size ($V^{AC114}_{32}$ = 12.8 meV, $V^{AC222}_{32}$  = 11.0 meV), coupling to excitons slowly decreases with size in ZZ GQDs ($V^{ZZ}_{32}$ = 9.7-21.0 meV).  These values are significantly larger compared to C132 and C132-L and explain the shorter S$_3$ lifetimes, indicating that geometry and asymmetry play a crucial role in minimizing coupling and relaxation rates.

The difference in relaxation rates between AC and ZZ GQDs arises due to large changes in exciton-phonon coupling.  Transitions in AC GQDs couple more strongly ($S_{\alpha\beta}>$ 0.01) to a range of phonon modes, including optical G and defect modes as well as hydrogen modes over 3000 cm$^{-1}$ (Figure \ref{hrf}(b)).  Strong electronic couplings over 10 meV also lead to the fast 10-100 fs lifetimes for $E/E_G >$ 1.5 across AC GQD sizes (indices 7-10 in Figure \ref{rates}(c)).  In contrast, the ZZ edge leads to extremely weak electron-phonon coupling, as shown in Figure \ref{hrf}(c).  For the S$_3$-S$_2$ transition, optical G and defect modes exhibit $S_{\alpha\beta} <$  0.01 even for the smallest dot.  This weak coupling to phonons reduces $F(\omega_{\alpha\beta})$ by orders of magnitude compared to the AC GQDs, leading to 20-40 ps lifetimes in ZZ GQDs.  This weak exciton-phonon coupling makes ZZ GQDs a promising material to generate long lifetimes if functionalization or geometries can be found to reduce the strong electronic coupling to excitons.

In contrast to AC GQDs, the S$_3$ lifetime decreases with increasing ZZ GQD size due to the fact that the electron-phonon coupling is already so weak in the smallest ZZ GQD.  As dot size increases, two competing effects determine $k_{\alpha\beta}$.  The transition energy decreases with size, requiring less phonons to induce the transition and increasing rates.  However, electron-phonon coupling also decreases with size, which reduces $S_{\alpha\beta}$ and $F(\omega_{\alpha\beta})$.  Whereas in AC GQDs the decrease of $\omega_{32}$ with size is balanced by reduced $S_{\alpha\beta}$, in ZZ GQDs the effect of a smaller transition energy dominates and leads to shorter carrier lifetimes.  In addition, $\omega_{32}$ decreases to 0.23 eV in ZZ216, matching the 0.20 eV (1600 cm$^{-1}$) optical G mode and creating a unique resonance condition that increases $F(\omega_{\alpha\beta})$ by an order of magnitude, leading to a single-phonon process and short 3.1 ps S$_3$ lifetime.  Further increasing ZZ GQD size would likely increase lifetimes because the resonance condition would no longer be satisfied and $V_{\alpha\beta}$ and $F(\omega_{\alpha\beta})$ would continue to decrease.

Our report of long excited carrier lifetimes in GQDs suggest that large transition energies and weak electronic coupling to excitonic states provides a new class of phonon bottleneck exploited in low-screening materials.  The bottleneck timescale is sensitive to the nanoscale structure of the GQD due to changes in electronic and electron-phonon coupling, providing a playground to tune state lifetimes by orders of magnitude for many applications.  Asymmetric GQDs and ligands (C132-L) minimize electronic coupling whereas ZZ edges minimize exciton-phonon coupling, both of which should be of interest for hot carrier extraction in photovoltaics.  Conversely, the rapid relaxation in AC GQDs could be beneficial in light-emitting applications.  Future work should explore functionalization and different GQD structures as a method to minimize coupling to excitons and maximize carrier lifetimes.

\begin{center}\bf{Methods}\end{center}

\textbf{Computational details.}  Electronic excitations in GQDs were obtained by solving for the twenty lowest eigenvalues of the Casida equation \cite{casida1995time} within linear response, time-dependent density functional theory (LR-TDDFT) \cite{runge1984density}.  All calculations have been completed using the Gaussian09 package \cite{frisch2009gaussian}, employing the B3LYP hybrid exchange-correlation functional within the adiabatic approximation and the 6-31G(d) Gaussian basis set.  The exact exchange contribution included in the hybrid functional reproduces the correct asymptotic behavior of the exchange potential that is necessary to accurately produce excitonic excitations below the electronic band gap in LR-TDDFT. Ground and excited state configurations were relaxed such that all atomic forces are less than 0.01 eV/$\textup{\AA}$.

Quantum molecular dynamics (QMD) simulations were performed to produce input structures for calculating the nonadiabatic coupling matrix elements ($V_{\alpha\beta}$) and room-temperature absorption spectra (see below).  Structures were initially relaxed at 0K and brought to 300K slowly through velocity rescaling.  After equilibrating at 300K using the canonical NVT ensemble, a 1 picosecond (ps) simulation was performed within the NVE ensemble with a 0.5 femtosecond (fs) time step.  Structures from the QMD simulation were then used as input to calculate LR-TDDFT excitations and nonadiabatic coupling using the B3LYP functional. 

\textbf{Absorption spectra.}  The zero-Kelvin absorption spectrum, $A^{0K}(\omega)$, for each GQD is calculated by weighting a delta function at each LR-TDDFT excitation energy, $E^{exc}_{0\alpha}=\hbar \omega_{0 \alpha}$, by its corresponding oscillator strength, $f_{0 \alpha}$:
\begin{equation}
\label{abs_noFCF}
A^{0K}(\omega) = \sum_{\alpha} f_{0 \alpha} \delta(\hbar \omega-\hbar \omega_{0\alpha}),
\end{equation}

\noindent where $f_{0 \alpha}=\frac{2 m_e \omega_{0\alpha}}{3 \hbar}|\langle \Psi_{0} | \textbf{r} | \Psi_{\alpha}\rangle |^2$, $|\Psi_{0} \rangle$ and $|\Psi_{\alpha} \rangle$ are the ground and excited state TDDFT wave functions,  $m_e$ is the electron mass, and $\langle \Psi_{0} | \textbf{r} | \Psi_{\alpha}\rangle$ is the transition dipole moment.  The delta function is modeled using a Gaussian function with 80 meV broadening.  

Forbidden transitions with near-zero oscillator strengths at 0K , known as dark states, can be activated by thermal motion at room temperature.  To better compare to experimental absorption spectra, we model these finite temperature effects for the C132-L GQD by calculating $\omega_{0\alpha}$ and $f_{0\alpha}$ over multiple time steps of a 300K QMD simulation.  The thermally averaged absorption spectra, $A(\omega)$, is then
\begin{equation}
A(\omega) = \frac{1}{N} \sum_{n=1}^N \sum_{\alpha} f^n_{0 \alpha} \delta(\hbar \omega-\hbar \omega^n_{0\alpha}),
\end{equation}

\noindent where $N$ is the total number of time steps in the MD run, and $\omega^n_{0\alpha}$ and $f^n_{0 \alpha}$ represent excitations and oscillator strengths at the $n$th time step.  The thermally averaged absorption spectrum is used for all results shown.

\textbf{Relaxation dynamics.}  We combine a procedure to calculate nonadiabatic coupling within LR-TDDFT \cite{tavernelli_exc_exc,tavernelli_ground_exc} with the reduced density matrix method \cite{may2008charge} to examine excited state population changes over time after an initial photoexcitation.  To the best of our knowledge, this is the first development of a computationally efficient method that can be used to investigate systems with hundreds of atoms while treating excited states rigorously using LR-TDDFT, which allows for a correct description of excitons.

Assuming a Markovian phonon bath and applying the secular approximation, the Redfield equations \cite{may2008charge} describe population changes in the reduced density matrix ($\sigma_{\alpha\beta}$) constructed from LR-TDDFT eigenstates as described by the transition rates $k_{\alpha\beta}$ (Eqn \ref{kab}).  The electronic nonadiabatic coupling ($V_{\alpha\beta}$) can be expressed as \cite{prezhdo1997evaluation}:

\begin{equation}
\label{vab}
V_{\alpha\beta}= \sum_k -\frac{i \hbar}{M_k} \langle \Psi_\alpha| \nabla_k| \Psi_\beta\rangle \langle \hat{p} \rangle,
\end{equation}

\noindent where $| \Psi_{\alpha} \rangle$ is the auxiliary wave function constructed using the Casida ansatz \cite{tavernelli_exc_exc}, $\nabla_k$ and $M_k$ are the nuclear gradient operator and mass for the $k$th atom, and $\langle \hat{p} \rangle$ is the expectation value of the momentum operator.  We evaluate this electronic coupling by running quantum molecular dynamics (QMD) simulations at 300K and calculating $V_{\alpha\beta}$ between excitations $\alpha$ and $\beta$ at each 0.5 fs time step using finite difference methods.  $V_{\alpha\beta}$ typically converges within 150 fs (Figure S3), highlighting the computational efficiency of the present method.

Assuming a quantum harmonic oscillator phonon bath, $F(\omega_{\alpha\beta})$ describes the density of phonon states coupled to a given nonadiabatic transition \cite{may2008charge}:
\begin{equation}
\label{fcdos_ft}
F(\omega_{\alpha\beta}) = \frac{1}{2\pi\hbar} \int_{-\infty}^{\infty} dt e^{i\omega_{\alpha\beta}t}e^{G(t)},
\end{equation}

\noindent where $G(t)=\sum_k S_{\alpha\beta}^k[(e^{-i\omega_k t} - 1)(n(\omega_k)+1) + (e^{i\omega_k t} - 1) n(\omega_k)]$.  $S^k_{\alpha\beta}$ is the Huang-Rhys factor, a dimensionless quantity representing the coupling strength of the $k$th normal mode to the transition, and $n(\omega_k)$ is the Bose-Einstein distribution for a phonon with energy $\hbar \omega_k$ at 300K.  $F(\omega_{\alpha\beta})$ can be expressed in terms of the nuclear wave function overlap and is connected to decoherence and the quantum Zeno effect \cite{kilina2013quantum}.

To examine the excited population as a function of time, we consider a system bathed in light with frequency $\omega$ at $t < 0$.   After the light is switched off at $t=0$, a nonequilibrium carrier distribution will relax due to interactions with phonons as described above.  The population change in $\sigma_{\alpha\beta}$ is then defined as $\Delta \eta(E,t)=\sum_{\alpha} \sigma_{\alpha\alpha}(t)\delta(E_{\alpha}-E) - \eta_{eq}$, where $E_{\alpha}$ is the energy of the $\alpha$th excitation and $\eta_{eq}$ is the equilibrium population.  Figures such as \ref{comp_sc_uc}(b) and \ref{ac_zz_rel}(a)-(c) plot $\Delta \eta(E,t)$ after the initial photoexcitation.

Please see the Supplementary Materials for more detailed explanation of the above methodology.

\begin{center}\bf{Acknowledgments}\end{center}
We acknolwedge support from the Department of Energy (Grant No. DOE/BES DE-FG02-02ER45995).   We thank NERSC for computing resources. 

\begin{center}\bf{Author contributions}\end{center}
J.P.T., I.H.C., and H.P.C. performed theoretical work.  J.P.T. developed code and conducted numerical calculations.  J.P.T., I.H.C., and H.P.C. all assisted in data analysis and interpretation.  J.P.T. wrote the paper with contributions from all authors.  H.P.C. supervised the research.

\begin{center}\bf{Additional information}\end{center}
Supplementary information is available in the online version of the paper.
Correspondence and requests for materials should be addressed to H.P.C. (cheng@qpt.ufl.edu).

\begin{center}\bf{Competing financial interests}\end{center}
The authors declare no competing financial interests.

\end{document}



\title{Supplementary Materials: Preventing rapid energy loss from electron-hole pairs to phonons in graphene quantum dots}


\author {J.P. Trinastic}
 \affiliation {Department of Physics and Quantum Theory Project, University of Florida, Gainesville, Florida, 32611, USA}
\author {Iek-Heng Chu}
\author {Hai-Ping Cheng*}
 \affiliation {Department of Physics and Quantum Theory Project, University of Florida, Gainesville, Florida, 32611, USA}


\date{\today}

\maketitle



\begin{center}\textbf{RELAXATION DYNAMICS METHODS}\end{center}
\textbf{Reduced Density Matrix Formalism}

To calculate transition rates between electronic excitations, we employ the reduced density matrix formalism that leads to the famous Redfield equations.  Comprehensive descriptions of the theory can be found in \cite{may2008charge} and \cite{toutounji_redfield}.  The formalism provides a computationally efficient method for calculating dissipative dynamics of a system embedded in a large, macroscopic environment.  Here, we will consider electrons to be the system of interest coupled to phonons from the surrounding lattice acting as the environment, or reservoir.

If the system and the reservoir make up a closed system, then we can write the total Hamiltonian as a sum of the system component, $\hat{H}_S$, reservoir component, $\hat{H}_R$, and the system-reservoir coupling component, $\hat{V}$:
\begin{equation}
\hat{H}_{tot}=\hat{H}_S+\hat{H}_R+\hat{V}.
\end{equation}

\noindent The system and bath can each be described by their respective eigenvalue equations:
\begin{equation}
\hat{H}_S|\Psi_{\alpha}\rangle=E_{\alpha}|\Psi_{\alpha}\rangle,
\end{equation}
\begin{equation}
\hat{H}_R|\phi_i\rangle=E_i|\phi_{i}\rangle,
\end{equation}

\noindent where $|\Psi_{\alpha}\rangle$ and $E_{\alpha}$ correspond to the system eigenstates and eigenvalues, and $|\phi_i\rangle$ and $E_i$ correspond to the reservoir eigenstates and eigenvalues.  The density operator corresponding to $\hat{H}_{tot}$ can be expressed as
\begin{equation}
\hat{\rho}(t) = \sum_n{w_n |\Phi_n\rangle\langle\Phi_n|},
\end{equation}

\noindent where $|\Phi_n\rangle$ is a pure state of the closed, combined system and $w_n$ is the corresponding probability.  However, we are only concerned with the system degrees of freedom, so we can define a reduced density operator (RDO), $\hat{\sigma}(t)$, by taking the trace over the reservoir space:
\begin{equation}
\hat{\sigma}(t) = Tr_R[\hat{\rho}(t)]=\sum_i{\langle \phi_i(t)|\hat{\rho}(t)|\phi_i(t)\rangle}.
\end{equation}

\noindent The RDO now only depends on the system's degrees of freedom and allows us to understand its dynamics due to coupling to the reservoir.  Moving to the interaction representation and assuming that the system-reservoir interaction is small enough that the system does not change the bath over time, the equation of motion for the RDO expanded to second order is
\begin{equation}
\label{eom_second_order}
\frac{d \sigma^I(t)}{dt}=-\frac{i}{\hbar}Tr_R[V^I(t),\sigma^I(0)\rho_B]-\frac{1}{\hbar^2}\int_0^t d\tau Tr_B [V^I(t),[V^I(t-\tau),\sigma^I(t-\tau)\rho_B]],
\end{equation}

\noindent where $\rho_B = \sum_i [e^{-\beta E_i}/(\sum_i e^{-\beta E_i})]$ represents the equilibrium thermal distribution of the reservoir (unchanged by the system) and $\beta = 1/k_BT$, where $k_B$ is Boltzmann's constant and $T$ is the temperature.  The first term on the righthand side of Eqn. \ref{eom_second_order} involves a thermal average of $V$, which will only lead to an overall shift of the Hamiltonian and does not introduce dissipative effects.  The second term contains information about the interaction between the system and reservoir that leads to irreversible energy dissipation from the system.  Therefore, we will neglect the first term in the following discussion.

We make two major assumptions to arrive at the final expression for relaxation dynamics between system eigenstates.  First, the second term on the righthand side of Eqn. \ref{eom_second_order} introduces memory effects, seen by the fact that $\sigma^I(t)$ depends on $\sigma^I(t-\tau)$.  If we assume that the bath dynamics are much faster than the system dynamics, then $\sigma^I(t)$ remains unchanged over the memory time scale and the effect can be ignored ($\tau=0$), known as the Markov approximation.  Applying this approximation and expressing the RDO in terms of the orthonormal system eigenstates, $|\Psi_{\alpha}\rangle$, we arrive at the Redfield equations:
\begin{equation}
\label{red1}
\frac{d\sigma_{\alpha\beta}^I(t)}{dt}=\sum_{\mu\nu}\sigma_{\mu\nu}^I(t)R_{\alpha\beta,\mu\nu}e^{i(\omega_{\alpha\beta}-\omega_{\mu\nu})t},
\end{equation}

\noindent where $\hbar\omega_{\alpha\beta}=E_{\alpha}-E_{\beta}$ is the transition energy between eigenstates.  $R_{\alpha\beta,\mu\nu}$ is a tetradic matrix that describes the rates of change of the elements of $\sigma^I(t)$.

The second approximation, known as the secular approximation, reduces the number of elements of $R_{\alpha\beta,\mu\nu}$ we need to calculate.  Examining Eqn. \ref{red1}, we can see that the righthand side depends on an exponential term that oscillates rapidly.  Over a long time average, the righthand side will vanish unless $\omega_{\alpha\beta}-\omega_{\mu\nu}=0$.  Imposing this constraint, the secular approximation reduces Eqn. \ref{red1} to two cases: (1) $\alpha=\beta,\mu=\nu$, corresponding to population transfer between eigenstates, and (2) $\alpha\neq \beta,\alpha=\mu,\beta=\nu$, corresponding to coherence dephasing that describes phase differences between eigenstates.  We can now write the Redfield equation for each of the two cases:

1) Population relaxation:
\begin{equation}
\label{red2}
\frac{d\sigma_{\alpha\alpha}^I(t)}{dt}=-\sum_{\mu}R_{\alpha\alpha,\mu\mu}\sigma^I_{\mu\mu}(t)
\end{equation}
\begin{equation}
\label{red3}
R_{\alpha\alpha,\mu\mu}=\delta_{\alpha\mu}\sum_{\kappa}k_{\alpha\kappa}-k_{\mu\alpha}
\end{equation}

2) Coherence dephasing:
\begin{equation}
\label{red4}
\frac{d\sigma_{\alpha\beta}^I(t)}{dt}=-R_{\alpha\beta,\alpha\beta}\sigma^I_{\alpha\beta}(t)
\end{equation}
\begin{equation}
\label{red5}
R_{\alpha\beta,\alpha\beta}=\sum_{\mu}\frac{1}{2}(k_{\alpha\mu}+k_{\beta\mu}) + \gamma_0,
\end{equation}

\noindent where $\gamma_0$ is a pure dephasing constant \cite{deph_constant}.  The rates $k_{\alpha\beta}$ can be shown to be \cite{tokmakoff_redfield,may2008charge}
\begin{equation}
\label{red_fgr}
k_{\alpha\beta}=\frac{2\pi}{\hbar}\sum_{i,j}f_i|\langle \phi_i \Psi_\alpha|V|\phi_j \Psi_\beta\rangle|^2\delta(\hbar \omega_{\alpha\beta}-\hbar \omega_{ij})
\end{equation}
\begin{equation}
\label{red_fgr_thermal}
=\frac{2\pi}{\hbar}\Big \langle \sum_{j}|\langle \phi_i \Psi_\alpha|V|\phi_j \Psi_\beta\rangle|^2\delta(\hbar \omega_{\alpha\beta}-\hbar \omega_{ij}) \Big \rangle,
\end{equation}

\noindent where $f_i=e^{-\beta E_i}/(\sum_i e^{-\beta E_i})$ and, in the second line, we have rewritten the initial distribution of quantum reservoir states as a thermal average over initial conditions, $\Big \langle ... \Big \rangle=\sum_i f_i$.  The sum over final reservoir states preserves conservation of energy.  The principle of detailed balance ensures probability conservation and provides the following relationship between forward and backward transition rates:
\begin{equation}
\label{det_bal}
k_{\alpha\beta}=e^{\hbar\omega_{\alpha\beta}/k_B T} k_{\beta\alpha}.
\end{equation}

\noindent In the present study, we are only concerned with population transfer between eigenstates of the system Hamiltonian, and therefore we only solve Eqn \ref{red2} and ignore coherence dephasing. 

In the next section, we describe the system-reservoir interaction $V$ that leads to nonadiabatic coupling between system eigenstates and determine how to calculate the matrix elements in Eqn. \ref{red_fgr}.

\noindent \textbf{Nonadiabatic Coupling}

Within the Born-Oppenheimer approximation, the total wave function is expressed as a product of electronic and nuclear components and the nuclear kinetic energy operator is neglected.  This leads to an electronic Hamiltonian that includes the electronic kinetic energy, electron-electron repulsive potential, and the electronic-nuclear attractive potential.  The eigenstates of the electronic Hamiltonian depend only parametrically on the nuclear coordinates.  Transitions between these adiabatic eigenstates can be described by including a nonadiabatic coupling perturbation defined by the nuclear kinetic energy operator, modeling transitions induced by lattice vibrations.  The matrix elements of this nonadiabatic coupling operator are
\begin{equation}
\label{nac_op_1}
\langle \phi_i \Psi_\alpha|V|\phi_j \Psi_\beta\rangle= V_{\alpha\beta,ij}=\langle \phi_i \Psi_\alpha|\sum_k -\frac{\hbar^2}{2M_k} \nabla^2_k | \phi_j \Psi_\beta\rangle,
\end{equation}

\noindent where $M_k$ is the nuclear mass and $k$ runs over the $K$ nuclear degrees of freedom, $| \phi_i \rangle = | \phi_i^1 \rangle | \phi_i^2 \rangle ... | \phi_i^K \rangle$.  If we only include terms to first order in the nuclear gradient operator, Eqn. \ref{nac_op_1} becomes
\begin{equation}
\label{nac_op_2}
V_{\alpha\beta,ij}=\sum_k -\frac{\hbar^2}{M_k}\langle \Psi_\alpha| \nabla_k| \Psi_\beta\rangle \langle \phi_i| \nabla_k| \phi_j \rangle,
\end{equation}

\noindent such that the operator becomes a product of electronic and nuclear coupling elements.  Previous work \cite{prezhdo1997evaluation} has shown that Eqn. \ref{nac_op_2} can be rewritten to good approximation as
\begin{equation}
V_{\alpha\beta,ij}\approx \sum_k -\frac{i \hbar}{M_k} \langle \Psi_\alpha| \nabla_k| \Psi_\beta\rangle \langle \hat{p} \rangle \prod_k \langle \phi_i^k | \phi_j^k \rangle,
\end{equation}

\noindent where $\langle \hat{p} \rangle = -i \hbar \langle \phi_i| \nabla_k| \phi_i \rangle$ is the expectation value of the nuclear momentum operator, which follows classical equations of motion according to Ehrenfest's theorem.  Using this expression to represent the matrix elements of the system-reservoir coupling $V$, our relaxation rate from Eqn. \ref{red_fgr_thermal} becomes 
\begin{equation}
\label{kab_nac}
k_{\alpha\beta}=\frac{2\pi}{\hbar}\Big \langle |\sum_k -\frac{i \hbar}{M_k} \langle \Psi_\alpha| \nabla_k| \Psi_\beta\rangle \langle \hat{p} \rangle|^2 \sum_{j} \prod_k | \langle \phi_i^k | \phi_j^k \rangle |^2 \delta(\hbar \omega_{\alpha\beta}-\hbar \omega_{ij}) \Big \rangle,
\end{equation}

\noindent where the electronic nonadiabatic coupling is $V_{\alpha\beta}=\sum_k -\frac{i \hbar}{M_k} \langle \Psi_\alpha| \nabla_k| \Psi_\beta\rangle \langle \hat{p} \rangle$.  Due to the order of magnitude difference in electron and nuclear relaxation times, we apply a decorrelation assumption to separate the thermal averaging of the electronic and nuclear components \cite{staib1995molecular,prezhdo1997evaluation}:
\begin{equation}
\label{kab_final}
k_{\alpha\beta}=\frac{2\pi}{\hbar}\Big \langle |V_{\alpha\beta}|^2 \Big \rangle \sum_{i,j} f_i \prod_k | \langle \phi_i^k | \phi_j^k \rangle |^2 \delta(\hbar \omega_{\alpha\beta}-\hbar \omega_{ij}),
\end{equation}

Thus, calculation of the transition rates between excitations reduces to the calculation of two major quantities: 1) the thermally averaged square of the electronic nonadiabatic coupling, $\Big \langle |V_{\alpha\beta}|^2 \Big \rangle$, and 2) the Franck-Condon-weighted phonon density of states (FC-DOS) , $F(\omega_{\alpha\beta})=\sum_{i,j} f_i \prod_k | \langle \phi_i^k | \phi_j^k \rangle |^2 \delta(\hbar \omega_{\beta \alpha}-\hbar \omega_{ji})$, the same term that appears in the absorption spectra but with $\omega = 0$.  Previous work has calculated transitions rates using Eqn. \ref{kab_final} but ignored the quantum nature of the nuclear wave functions, essentially setting the nuclear overlap to one \cite{kilin2010relaxation,chu2013first}:
\begin{equation}
\label{kab_nofcdos}
k_{\alpha\beta}=\frac{2\pi}{\hbar}\Big \langle |V_{\alpha\beta}|^2 \Big \rangle \sum_{i,j} f_i \delta(\hbar \omega_{\alpha\beta}-\hbar \omega_{ij}).
\end{equation}

\noindent This approximation has given results matching experiment in some cases, however likely overestimates relaxation rates because it weights all phonon modes equally.  Especially in systems with a large number of bath modes and weak electron-phonon coupling, it is important to account for the nuclear overlap to only include modes coupled to a given transition.  

\noindent \textbf{Nonadiabatic Coupling within LR-TDDFT}

We calculate the electronic nonadiabatic coupling matrix elements using the auxiliary many-body wave function originally proposed by Casida  \cite{casida2009time} and Tavernelli \cite{tavernelli2009nonadiabatic} to describe excited state wave functions within LR-TDDFT.  In this case, wave functions are expressed as linear combinations of singly excited Slater determinants of Kohn-Sham orbitals, determined from the solution of the Casida equation:
\begin{equation}
\label{casida_ansatz}
| \Psi_{\alpha} \rangle = \sum_{ma}(X_{\alpha}^{ma}+Y_{\alpha}^{ma}) \hat{a}^{\dagger}_a \hat{a}_m | \Psi_{0} \rangle = \sum_{ma}(X_{\alpha}^{ma}+Y_{\alpha}^{ma}) | \Psi_{m -> a} \rangle,
\end{equation}

\noindent where $m$ and $a$ index occupied and unoccupied Kohn-Sham orbitals, respectively, $\hat{a}^{\dagger}_a$ and $\hat{a}_m$ are creation and annihilation operators, $X_{\alpha}^{ma}$ and $Y_{\alpha}^{ma}$ are the normalized coefficients corresponding to an excitation from the $m$th to $a$th orbital and de-excitation from the $a$th to $m$th orbital, respectively, and $ | \Psi_{0} \rangle$ is the ground state Slater determinant. This auxiliary wave function has been shown to give the exact first-order nonadiabatic coupling between the ground state and any excited state \cite{tavernelli2009nonadiabatic} and is an accurate approximation to the first-order coupling between excited states \cite{tavernelli_exc_exc}.  The $V_{\alpha\beta}$ matrix elements can be calculated by converting the nuclear gradient operator to a time derivative and using the finite difference method between adjacent timesteps in a quantum molecular dynamics simulation:
\begin{eqnarray}
\label{timeder_new}
V_{\alpha\beta} = \sum_k -\frac{i \hbar}{M_k} \langle \Psi_\alpha| \nabla_k| \Psi_\beta\rangle \langle \hat{p} \rangle \\
\nonumber = -i \hbar \langle \Psi_\alpha| \frac{\partial}{\partial t} | \Psi_\beta\rangle \\
\nonumber \approx - \frac{i \hbar}{2 \Delta t} [\langle \Psi_\alpha(t)|\Psi_\beta (t+\Delta t)\rangle - \langle \Psi_\alpha(t + \Delta t)|\Psi_\beta (t)\rangle],
\end{eqnarray}

\noindent where $\Delta t$ is the 0.5 fs time step of the molecular dynamics simulation.  At each time step, we calculate the nonadiabatic coupling between the many-body wave functions using Eqns. \eqref{casida_ansatz} and \eqref{timeder_new} .  In particular, we can decompose the many-body wave function overlap from the finite-difference method in Eqn. \eqref{timeder_new} as a sum over singly excited Slater determinant overlaps:
\begin{eqnarray}
\label{sd_ovlap}
\langle \Psi_\alpha(t)|\Psi_\beta (t+\Delta t)\rangle = \\
\nonumber \sum_{ma}\sum_{nb}(X_{\alpha}^{ma}(t)+Y_{\alpha}^{ma}(t))(X_{\beta}^{nb}(t+ \Delta t)+Y_{\beta}^{nb}(t + \Delta t)) \langle \Psi_{m -> a}(t)| \Psi_{n -> b} (t+ \Delta t) \rangle.
\end{eqnarray}

\noindent The singly excited Slater determinant overlap in Eqn. \eqref{sd_ovlap} can be decomposed using Slater-Condon rules and calculated using the overlaps between single-particle Kohn-Sham orbitals making up each determinant.  These overlaps are nonzero when either the determinants are the same or differ by only one single-particle orbital.  Thus, the nonadiabatic coupling matrix elements ($V_{\alpha\beta}$) are found by weighting the Kohn-Sham orbital overlaps by the corresponding LR-TDDFT excitation coefficients ($X_{\alpha}^{ma}$ and $Y_{\alpha}^{ma}$) and summing over all single-particle transitions that make up each many-body excitation.

The thermally averaged electronic nonadiabatic coupling that enters into Eqn. \eqref{kab_final} is calculated by averaging the value of $V_{\alpha\beta}$ calculated between adjacent time steps across a 300K QMD trajectory consisting of $N$ time steps:
\begin{equation}
\label{vab_conv_time}
\Big \langle |V_{\alpha\beta}|^2 \Big \rangle = \frac{1}{N} \sum_{i=1}^N |V_{\alpha\beta}|^2.
\end{equation}

\noindent \textbf{Franck-Condon-Weighted Density of States}

We calculate the Franck-Condon-weighted density of states assuming that the phonon reservoir can be described as a harmonic bath and that the ground state normal modes are an accurate approximation for all excited state modes.  In this case, nuclear wave functions can be expressed as eigenstates of a quantum harmonic oscillator and $F(\omega_{\alpha\beta})$ can be rewritten as \cite{may2008charge}:
\begin{equation}
\label{fcdos_ft}
F(\omega_{\alpha\beta}) = \frac{1}{2\pi\hbar} \int_{-\infty}^{\infty} dt e^{i\omega_{\alpha\beta}t}e^{G(t)}
\end{equation}

\noindent and
\begin{equation}
\label{g_t}
G(t)=\sum_k S_{\alpha\beta}^k[(e^{-i\omega_k t} - 1)(n(\omega_k)+1) + (e^{i\omega_k t} - 1) n(\omega_k)],
\end{equation}

\noindent where $\omega_k$ is the frequency of the $k$th mode, $t$ is time, and $n(\omega_k) = (e^{\hbar \omega_k/k_B T} - 1 )^{-1}$ is the Bose-Einstein distribution that provides temperature dependence.  $S^k_{\alpha\beta}$ is the Huang-Rhys factor, defined as
\begin{equation}
\label{hrfactor}
S^k_{\alpha\beta} = \frac{\mu_k \omega_k}{\hbar}d_k^2,
\end{equation}

\noindent where $\mu_k$ is the reduced mass of the $k$th normal mode and $d_k$ is the projection of the mass-weighted, configurational difference between excitations $\alpha$ and $\beta$ onto the $k$th mass-weighted, normal mode eigenvector.  The Huang-Rhys factor measures the coupling strength between a given electronic transition and the $k$th mode and is the primary contribution to altering the FC-DOS compared to the complete phonon DOS.  Modes that couple strongly with a given transition corresponds to a larger peak at that mode energy in the FC-DOS.  Multiphonon processes are also included in Eqn. \ref{fcdos_ft} that result in diminished peaks at multiples of a given phonon energy.  Finally, the reorganization energy, $\lambda$, for a given electronic transition is defined as
\begin{equation}
\label{reorg_def}
\lambda_{\alpha\beta} = \sum_k S^k_{\alpha\beta} \hbar \omega_{\alpha\beta},
\end{equation}

\noindent which measures the energy dissipated in the system to relax to its minimum after a vertical electronic transition.  Small reorganization energies correspond to transitions between states with similar geometries.

In summary, the calculation of phonon-assisted relaxation between TDDFT excitations requires three ingredients: 1) the reduced density matrix constructed from the eigenstates of the Casida equation within linear-response TDDFT; 2) thermally averaged nonadiabatic couplings $V_{\alpha\beta}$ calculated along a QMD trajectory within the LR-TDDFT framework; and 3) the Franck-Condon-weighted density of states (FC-DOS) calculated within the harmonic approximation and requiring the ground state normal modes and excited state configurations of each system.  


\begin{figure}[h]
  \centering
    \includegraphics[scale=2]{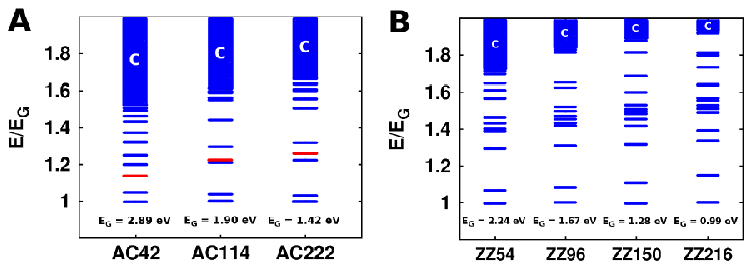}
    \caption{LR-TDDFT excitations for various GQD types and sizes.  a) Excitation levels for armchair (AC) GQDs. b) Excitation levels for zigzag (ZZ) GQDs.  Excitation energy is given as a function of $E/E_G$, where $E_G$ is the lowest LR-TDDFT excitation energy.  Blue horizontal lines indicate excitation energies.  The red excitation indicates a state of interest that changes its position relative to other states as a function of size in AC GQDs.  $C$ stands fo the continuum level, at which point transition energies between states are only several meV.}
  \label{ac_zz_exc}
\end{figure}

\begin{figure}[h]
  \centering
    \includegraphics[scale=1]{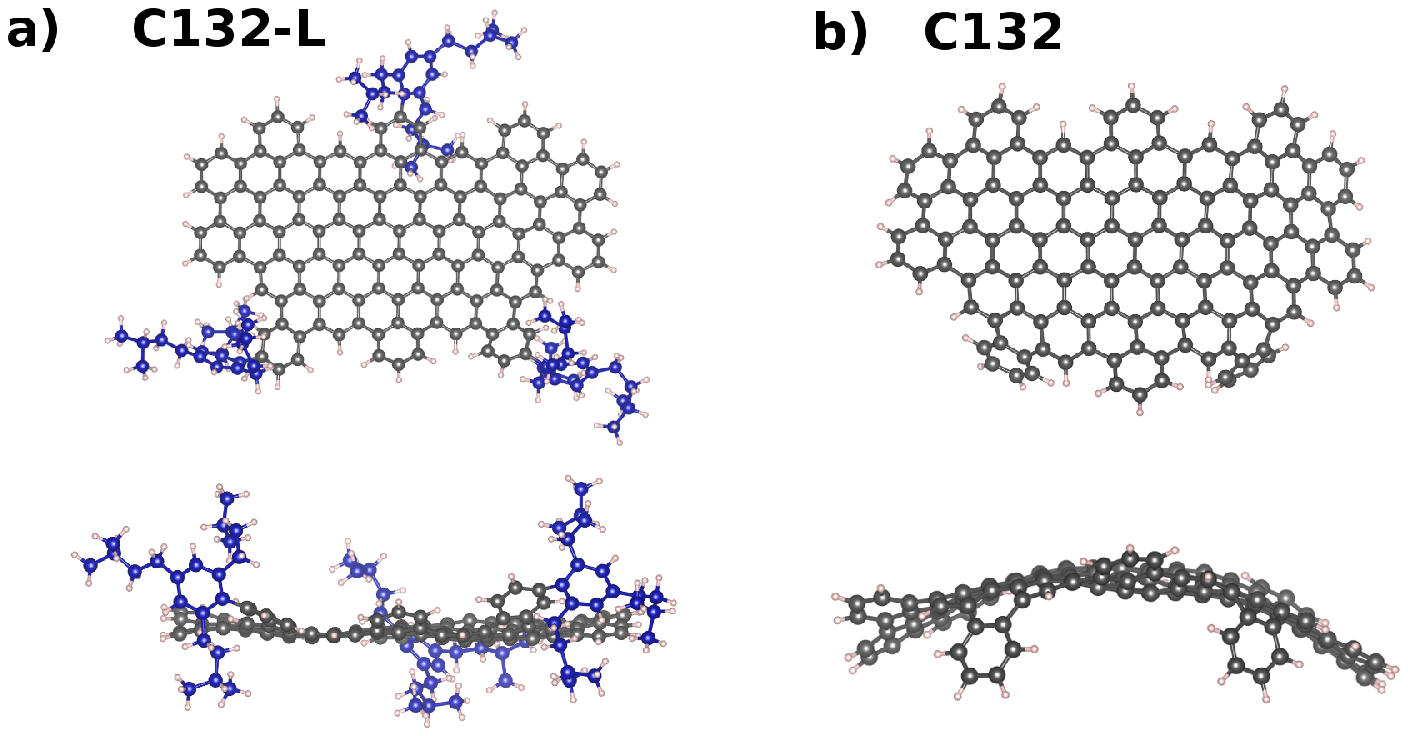}
    \caption{Configuration of a) C132-L and b) C132 GQD at representative timestep in the 300K molecular dynamics trajectory.  Medium gray spheres are carbon atoms in the body of the GQD, medium blue spheres are carbon atoms in the ligands, and small pink spheres are hydrogen atoms.}
  \label{sc_uc_300}
\end{figure}

\begin{figure}[h]
  \centering
    \includegraphics[scale=1]{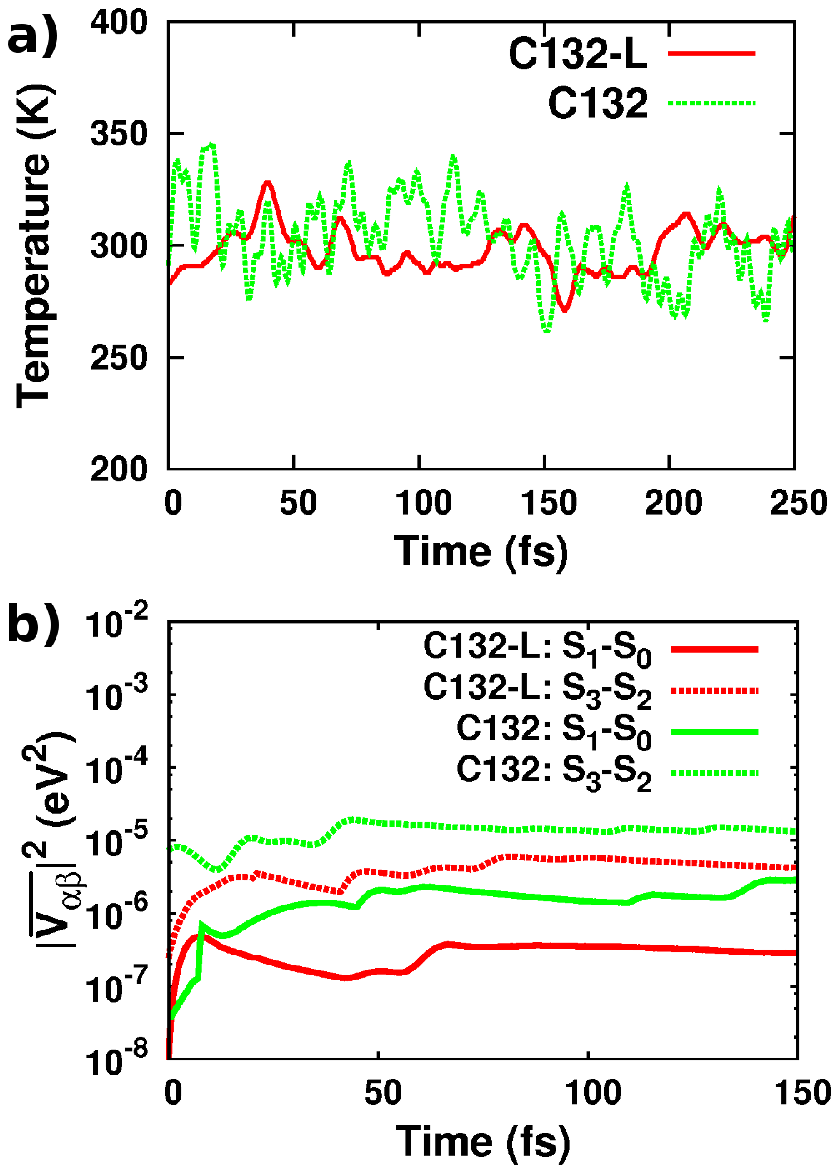}
    \caption{a) Temperature and b) average square of the nonadiabatic coupling ($<|V_{\alpha\beta}|^2>$) as a function of time from the 300K $\textit{ab initio}$ molecular dynamics simulation run within the microcanonical (NVE) ensemble for the C132-L and C132 GQDs.}
  \label{temp_vac}
\end{figure}

\begin{table}
   \caption{Vertical excitation energies ($E^{exc}_{0\alpha}$) and oscillator strengths ($f_{0\alpha}$) for low-lying states at 0K for armchair (AC) GQDs.}\label{ac_exc}
\resizebox{\columnwidth}{!}{%
\begin{tabular}{l*{7}{c}r}
& AC42 & &  AC114 & & AC222  &  & 	 \\
Excitation & $E^{exc}_{0\alpha}$ (eV) & $f_{0\alpha}$  & $E^{exc}_{0\alpha}$ (eV) & $f_{0\alpha}$ &  $E^{exc}_{0\alpha}$ (eV) & $f_{0\alpha}$\\
\hline
$\textup{S}_0 \to \textup{S}_\textup{1}$    & 2.894 & 0.000 & 1.900 & 0.000 & 1.418 & 0.000  \\
$\textup{S}_0 \to \textup{S}_\textup{2}$    & 3.028 & 0.000 & 1.970 & 0.000 & 1.462 & 0.000  \\
$\textup{S}_0 \to \textup{S}_\textup{3}$ & 3.291 & 0.000 & 2.306 & 1.328 & 1.738 & 2.029  \\
$\textup{S}_0 \to \textup{S}_\textup{4}$ & 3.476 & 0.722 & 2.309 & 0.000 & 1.783 & 0.000  \\
$\textup{S}_0 \to \textup{S}_\textup{5}$ & 3.624 & 0.000 & 2.462 & 0.000 & 1.873 & 0.000  \\
$\textup{S}_0 \to \textup{S}_\textup{6}$    & 3.834 & 0.000 & 2.739 & 0.000 & 2.139 & 0.000  \\
\hline
\end{tabular}}
\label{ac_exc}
\end{table}

\begin{table}
   \caption{Vertical excitation energies ($E^{exc}_{0\alpha}$) and oscillator strengths ($f_{0\alpha}$) for low-lying states at 0K for zigzag (ZZ) GQDs.}\label{zz_exc}
\resizebox{\columnwidth}{!}{%
\begin{tabular}{l*{8}{c}r}
& ZZ54 & &  ZZ96 & & ZZ150 &  & ZZ216 &  &	 \\
Excitation & $E^{exc}_{0\alpha}$ (eV) & $f_{0\alpha}$  & $E^{exc}_{0\alpha}$ (eV) & $f_{0\alpha}$ &  $E^{exc}_{0\alpha}$ (eV) & $f_{0\alpha}$  &  $E^{exc}_{0\alpha}$ (eV) & $f_{0\alpha}$  \\
\hline
$\textup{S}_0 \to \textup{S}_\textup{1}$     & 2.239 & 0.000 & 1.667 & 0.000 & 1.280 & 0.000 & 0.992 & 0.000  \\
$\textup{S}_0 \to \textup{S}_\textup{2}$     & 2.391 & 0.000 & 1.806 & 0.000 & 1.422 & 0.000 & 1.139 & 0.000 \\
$\textup{S}_0 \to \textup{S}_\textup{3}$ & 2.905 & 0.929 & 2.179 & 1.143 & 1.688 & 1.279 & 1.324 & 1.311  \\
$\textup{S}_0 \to \textup{S}_\textup{4}$ & 3.122 & 0.000 & 2.367 & 0.000 & 1.818 & 0.000 & 1.379 & 0.000 \\
$\textup{S}_0 \to \textup{S}_\textup{5}$    & 3.140 & 0.000 & 2.386 & 0.000 & 1.865 & 0.000 & 1.474 & 0.000 \\
$\textup{S}_0 \to \textup{S}_\textup{6}$    & 3.205 & 0.000 & 2.429 & 0.000 & 1.905 & 0.000 & 1.497 & 0.000 \\
\hline
\end{tabular}}
\label{zz_exc}
\end{table}

\begin{table}
\caption{Transition energy ($\omega_{\alpha\beta}$), electronic nonadiabatic coupling ($V_{\alpha\beta}$), and Franck-Condon-weighted density of states ($F(\omega_{\alpha\beta})$) for sample transitions between LR-TDDFT excitations in C132-L and C132 GQDs.}\label{table_reldata_c132}
\resizebox{\columnwidth}{!}{%
\begin{tabular}{l*{7}{c}r}
Transition & $\omega_{\alpha\beta}$ (eV) & &  $V_{\alpha\beta}$ (meV) & & $F(\omega_{\alpha\beta})$ (1/eV) &  \\
                & C132-L & C132  & C132-L & C132&  C132-L & C132 &  \\
\hline
$\textup{S}_\textup{1}$-$\textup{S}_\textup{0}$ & 1.918 & 1.920 & 0.794 & 2.550 & 0.073 & 0.052 \\
\hline
$\textup{S}_\textup{2}$-$\textup{S}_\textup{1}$ & 0.060 & 0.061 & 20.075 & 36.606 & 0.139 & 0.044 \\
\hline
$\textup{S}_\textup{3}$-$\textup{S}_\textup{2}$ & 0.295 & 0.302 & 1.649 & 2.324 & 0.134 & 0.065 \\
$\textup{S}_\textup{3}$-$\textup{S}_\textup{1}$ & 0.355 & 0.363 & 1.729 & 5.206 & 0.098 & 0.027 \\
\hline
$\textup{S}_\textup{4}$-$\textup{S}_\textup{3}$ & 0.069 & 0.062 & 16.733 & 8.087 & 0.270 & 0.168 \\
$\textup{S}_\textup{4}$-$\textup{S}_\textup{2}$ & 0.364 & 0.364 & 1.691 & 3.701 & 0.194 & 0.089 \\
$\textup{S}_\textup{4}$-$\textup{S}_\textup{1}$ & 0.424 & 0.425 & 1.905 & 7.183 & 0.110 & 0.034 \\
\hline
$\textup{S}_\textup{9}$-$\textup{S}_\textup{8}$ & 0.071 & 0.076 & 2.156 & 9.105 & 2.198 & 1.617 \\
\hline
\end{tabular}}
\label{c132_tab}
\end{table}

